\documentclass[10pt,twocolumn,aps,pre,superscriptaddress]{revtex4-1} %

\usepackage{graphicx}
\usepackage{dcolumn}
\usepackage{amsmath}
\usepackage{amssymb}
\usepackage{bm}
\usepackage{color}
\usepackage{mathrsfs}
\usepackage{multirow}
\usepackage{ulem}

\begin{document}

\title{Clustering of Self-Propelled Triangles with Surface Roughness}

\author{Sven Erik Ilse}
\affiliation{Institute for Computational Physics, University of Stuttgart, Allmandring 3, 70569 Stuttgart, Germany}

\author{Christian Holm}
\affiliation{Institute for Computational Physics, University of Stuttgart, Allmandring 3, 70569 Stuttgart, Germany}

\author{Joost de Graaf}
\email{jgraaf@icp.uni-stuttgart.de}
\affiliation{Institute for Computational Physics, University of Stuttgart, Allmandring 3, 70569 Stuttgart, Germany}

\date{\today}
\
\begin{abstract}
Self-propelled particles can spontaneously form dense phases from a dilute suspension in a process referred to as motility-induced phase separation. The properties of the out-of-equilibrium structures that are formed are governed by the specifics of the particle interactions and the strength of the activity. Thus far, most studies into the formation of these structures have focused on spherical colloids, dumbbells, and rod-like particles endowed with various interaction potentials. Only a few studies have examined the collective behavior of more complex particle shapes. Here, we increase the geometric complexity and use Molecular Dynamics simulations to consider the structures formed by triangular self-propelled particles with surface roughness. These triangles either move towards their apex or towards their base,~\textit{i.e.}, they possess a polarity. We find that apex-directed triangles cluster more readily, more stably, and have a smoother cluster interface than their base-directed counterparts. A difference between the two polarities is in line with the results of [H.H. Wensink,~\textit{et al.}, Phys. Rev. E~\textbf{89}, 010302(R) (2014)], however, we obtain the reversed result when it comes to clustering, namely that apex-directed particles cluster more readily. We further show that reducing the surface roughness negatively impacts the stability of the base-directed structures, suggesting that their formation is in large part due to surface roughness. Our results lay a solid foundation for future experimental and computational studies into the effect of roughness on the collective dynamics of swimmers.
\end{abstract}

\maketitle

\section{\label{sec:intro}Introduction}

Active matter is a state of matter in which the constituents constantly consume energy to perform work: the system is out of equilibrium~\cite{ramaswamy10a,marchetti13a}. Over the past decade, there has been a quantum leap in the ability to fabricate artificial self-propelled particles with colloidal length scales (1~nm to 1~$\mu$m)~\cite{paxton04a,wang06a,howse07a,valadares10a,ebbens12a,lee14a,brown14a,ebbens14a,simmchen14a}. This has led to a strong experimental interest in model systems comprised of such particles~\cite{ebbens10a,hong10a,sengupta12a,wang13a,sanchez15a}, as insights into the workings of fundamentally out-of-equilibrium processes may be gleaned from them~\cite{cates12a,cates15a}. Moreover, these man-made active systems display dynamics that are reminiscent of the patterns created by living organisms~\cite{marchetti13a}. Such organisms include biological swimmers which have a colloidal length scale,~\textit{e.g.}, bacteria~\cite{Sokolov07,Schwarz-Linek12,Reufer14}, algae~\cite{Polin09,Geyer13}, and sperm~\cite{Woolley03,Riedel05,Ma14}. The study of man-made self-propelled particles is therefore envisioned to lead to better understanding of the complexities of life itself. Finally, the level of control over the synthesis of active particles has resulted in a wide variety of particle shapes being available for these types of investigation. From relatively --- perhaps deceptively~\cite{brown14a,ebbens14a,brown15a-pre} --- simple hemispherical Janus particles~\cite{ebbens12a,howse07a,simmchen14a}, to dumbbells~\cite{valadares10a,wang14d}, hollow cones~\cite{Solovev09,Mei11}, L-shaped particles~\cite{kummel13,tenhagen14}, stomatocytes~\cite{nijs2013}, and many others. This and the diversity of swimmer shapes found in nature, leads naturally to the question: To what extend does shape influence the macroscale behavior in suspensions of these particles?

The exciting opportunities that active matter --- both biological and man-made systems --- presents for the understanding of the organization of life, as well as the nature of out-of-equilibrium systems, have piqued the curiosity of the simulation and theory communities. Simulations and theory offer the ability to disentangle the complexities of the self-propulsion mechanism and its influence on the collective dynamics of particles from the other particle and system properties. One of the first models for self-propelled particles was introduced 1995 by Vicsek~\textit{et al.}~\cite{vicsek95}, who considered the formation of aligned bands of point-like swimmers with only a local orientation-aligning potential. Since then, many different swimmer models have been considered that go beyond the point approximation. Particularly popular is the active Brownian particle model~\cite{zheng13,Stenhammar13}, which simulates self-propelled spherical particles and makes minimal assumptions for the driving mechanism. The mechanism is a simple persistent motion due to a constant force, which is applied in a fixed direction in the frame co-moving with the particle. For this model a multitude of interaction potentials has been studied and the state diagram of suspensions of these particles was found to be strongly influenced by the choice of interaction~\cite{Redner13}.

The first simulations to consider shape anisotropy in the context of collective motion focused on simple dumbbell~\cite{Thakur12,gonnella14,Cugliandolo15,tung16} and rod-shaped~\cite{Peruani06,wensink08,Ginelli10,Yang10,Abkenar13} swimmers. These models more accurately represent the oblong body of biological swimmers~\cite{Sokolov07,Schwarz-Linek12,Reufer14,Polin09,Geyer13,Woolley03,Riedel05,Ma14}, as well as the shape of catalytically propelled Au-Pt nanorods~\cite{paxton04a,wang06a} and hollow cones~\cite{Solovev09,Mei11}. Simulations of rigid rods and flexible chains~\cite{isele15} provide a good opportunity to assess the influence of shape, when compared to results for spherical active Brownian particles. In particular, it has become clear that shape can strongly enrich the state diagram, in much the same way as it does for the passive counterpart. That is, passive hard spheres have a phase diagram with a liquid and crystalline phase~\cite{hansen90}, while rod-like particles can also form an isotropic and smectic phase~\cite{frenkel88,bolhuis97}. Similarly, the state diagram of active hard spheres --- at low density --- displays an isotropic state and a state in which dense clusters form amid a dilute gas of active particles~\cite{Stenhammar13}; these clusters may also exhibit cooperative motion. Dynamic clusters form in a process that is referred to as motility-induced phase separation (MIPS). The state diagram for self-propelled rods is much more complex, as such particles can also display swarming, turbulence, and laning~\cite{wensink12}.

Only a few studies have ventured beyond simple rod and dumbbell shapes. Wensink~\textit{et al.}~\cite{wensink14}, for example, studied the behavior of sperm-type and algae-like swimmers. These authors approximated such particles by a ``snow-cone-like'' assembly of spheres that has a polarity,~\textit{i.e.}, the orientation of the self-propulsive force with respect to the shape anisotropy. In addition, they considered crescent-shaped bacteria, also possessing polarity induced by their direction of motion. Very recently, Mallory and Cacciuto considered triangular particles with selectively attractive edges, in order to demonstrate how self-propulsion can be improve the yield of hexagonal capsids~\cite{mallory16}. Finally, there is the work by Prymidis~\textit{et al.}~\cite{prymidis16}, which considers self-propelled squares. These studies indicate that there is a strong influence of shape on the clustering and dynamics of self-propelled particles. However, in spite of these recent efforts, there remains a large gap between this type of study into the effect of geometry and the next level in complexity, which attempts to accurately model specific microswimmers, such as \textit{Trypanosomes}~\cite{alizadehrad15}, sperm~\cite{Elgeti10}, and \textit{E. coli}~\cite{Hu15}.

\begin{figure}[!htb]
\begin{center}
\includegraphics[scale=1.0]{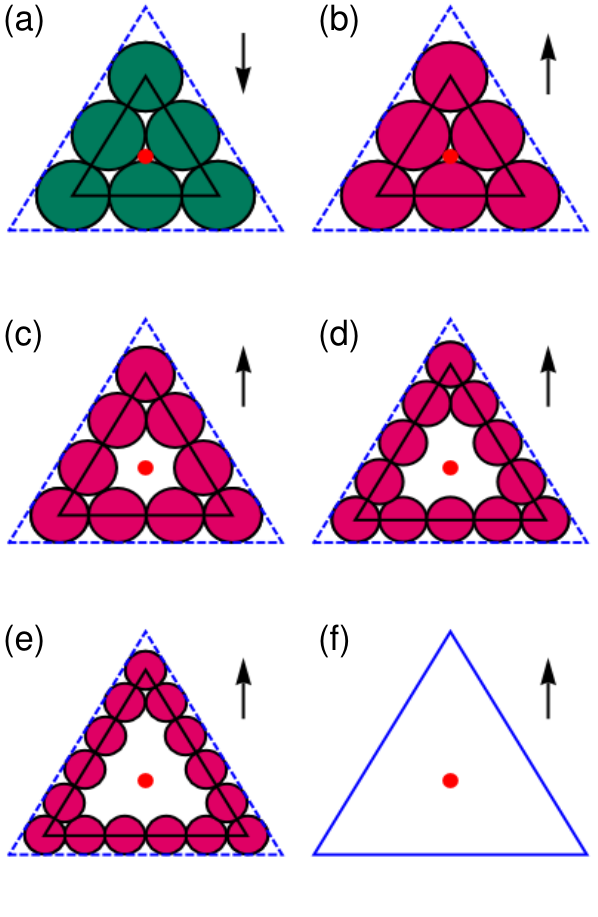}
\end{center}
\caption{\label{fig:model}2D representations of our rough equilateral triangle model. The model consists of several molecular dynamics (MD) beads (colored disks) that form the edge of the triangle (blue dashed line) and are connected rigidly to a central bead (red dot). The arrow shows the direction in which the triangle self-propels. (a) A base-directed (BD) triangle with $N=3$ MD beads per edge. (b) An apex-directed (AD) triangle with $N=3$. (c-e) AD triangles with $N=4$, $5$, and $6$ beads per edge, respectively. (f) The limiting AD triangle for $N\uparrow\infty$ (blue solid line). The centers of the edge MD beads are located on a smaller equilateral triangle (black line), which has edge length $L'$, the limiting triangle (blue) has edge length $L$.}
\end{figure}

In this paper, we therefore aim to extend this intermediate regime and consider a level of geometric complexity that goes beyond the rod-like, namely a triangular swimmer (without attractive edge decoration). This is the missing step in the series: disk, rod, $\dots$, square, etc. We study a quasi-two-dimensional (quasi-2D) system of active equilateral triangles moving either towards their apex or towards their base. This is similar to the polar swimmers considered by Wensink~\textit{et al.}~\cite{wensink14}, but we restrict ourselves here to the structure formation in pure phases of triangular particles. Our system is a variation of the active Brownian model: the particles experience translational and rotational Brownian motion in addition to a persistence force that leads them to self-propel. The triangles are comprised of spheres that make up the edge and endow it with a roughness, see Fig.~\ref{fig:model}, which provides us with a second variational parameter, beyond the polarity. Increasing the number of spheres used to discretize the triangle edges leads to a decrease of this roughness. Such roughness is a common and overlooked feature of a large set of similarly constructed simulation models (albeit with low surface roughness), see,~\textit{e.g.}, Refs.~\cite{wensink12,kaiser13,wensink14,mallory16}, and has not been investigated systematically thus far.

We construct state diagrams for both apex-directed (AD) and base-directed (BD) active triangles by varying the density and P{\'e}clet ($Pe$) number, which gives the ratio of self-propulsion to translational diffusion. We find MIPS for both systems, but the nature of this phase separation differs for the two models. AD triangles tend to phase separate at intermediate P{\'e}clet numbers and densities, forming large clusters that become system-spanning at sufficiently high values of the density and $Pe$. However, BD triangles are less inclined to form system spanning clusters. Large clusters of BD triangles have a rougher interfacial structure than their AD counterparts and are more likely to periodically break up. Both types of triangle also form a variety of small short-lived clusters at lower densities and $Pe$. These oligomers display both active translation and rotation, depending on the way the constituents have self-assembled, and give insight into the structure and stability of the system-spanning clusters. 

Our findings have an asymmetry between polar (AD in our terms) and antipolar (BD in our terms) particles, similar to the one observed by Wensink~\textit{et al.}~\cite{wensink14}. However, their antipolar particles tend to form clusters, whereas their polar particles do not and form large-scale swarms instead. We discuss the possible origin of this inversion with respect to our triangular system. We further show that the stability of the BD-triangle clusters is reduced by decreasing the triangle surface roughness. This suggests that in the flat-edged limit, it may be the case that only the AD system can form large stable clusters. In this context we also comment on the similarities between our results and the observations of Prymidis~\textit{et al.}~\cite{prymidis16}, who found oscillating phases of self-propelled squares. Finally, we consider the relevance of surface-roughness in physical systems. Here, we differentiate between ``dry'' (granular) active matter, which we simulate, and wet active matter, for which hydrodynamic interactions are taken into account.

The results presented in this paper underpin the sensitive nature of the clustering of self-propelled particles on the specifics of the particle shape and the polarity. In addition, we show that surface roughness can be a crucial parameter in stabilizing the MIPS. This study will form the basis for future work on the effects of hydrodynamic interactions on the behavior of these active triangles, as well as studies into the collective dynamics other polygonal swimmers with surface roughness and further investigation of the effect of roughness in general.

\section{\label{sec:methods}Methods}

In this section we present the construction of our triangular swimmers. This is followed by a listing and justification of the specific parameters that were chosen for our simulations. Finally, we detail the methods by which the MIPS was analyzed.

\subsection{\label{sub:triangle}Triangular Swimmers}

We construct triangular swimmers by placing spherical molecular dynamics (MD) beads equidistantly onto the edges of an equilateral triangle with edge length $L'$ and connect these rigidly to the triangle's center, see the black triangle and red dot in Fig.~\ref{fig:model}, respectively. These MD beads interact with each other via a Weeks-Chandler-Anderson (WCA) potential, which is specified by
\begin{align}
  \label{eq:WCA} U_{\mathrm{WCA}}(r) &= \left\{ \begin{array}{rc} 4 \epsilon \left[ \left( \frac{\sigma}{r} \right)^{12} - \left( \frac{\sigma}{r} \right)^{6} + \frac{1}{4} \right], & r \le 2^{1/6} \sigma \\[1.5em] 0, & r > 2^{1/6} \sigma \end{array} \right. ,
\end{align}
where $r$ is the inter-particle distance, $\sigma$ is the ``diameter'' of the MD bead, and $\epsilon$ is the interaction strength. The WCA potential models an excluded-volume interaction between the MD beads. We choose $\sigma$ in such a way that $L' = 2^{1/6} \sigma N$, with $N$ the numbers of beads per edge. Henceforth, we will refer to $N$ as a roughness parameter. In the limit $N \uparrow \infty$ the diameter $\sigma \downarrow 0$, leading to an increasingly smooth triangle. The limiting flat-faced triangle has an edge length of $L$, see Fig.~\ref{fig:model}f. For any finite value of $N$ the value of $L'$ is chosen such that the edge of the MD beads' WCA interaction range ($2^{1/6}\sigma$) touches the triangle with edge length $L$, see Fig.~\ref{fig:model}a-e. We made this choice to ensure that the convex hull of the triangles that we simulate converges to the limiting equilateral triangle in a monotonic fashion.

Note that our choice to employ spherical MD beads implies that the triangle models are three-dimensional (3D) objects, rather than two-dimensional (2D) constructs. For any finite value of $\sigma$ the models, in fact, represent triangular prisms. However, since we constrain these triangles to move in a plane, as we will come back to shortly, the third dimension is not relevant to the discussion. We therefore choose to present our results using 2D terms such as ``triangles'' and ``area'' instead --- though the reader should be aware that all our simulations are in principle 3D, with a quasi-2D constraint imposed. 

Before we turn to the self-propulsion of the triangles, we should also briefly comment on the units that we employ throughout. In the remainder of the text, we employ a unit-free notation. That is, we dedimensionalize lengths by a length unit, times by a time unit, etc. The length unit we employ is $\ell = \sigma_{(N=3)} \equiv 1$,~\textit{i.e.}, the diameter of the MD beads for a discretization of $N=3$ beads per edge length. This choice makes it that we can, for example, write $L = 2^{1/6}(2+\sqrt{3}) \approx 4.19$ for the edge length of the limiting triangle and $\ell^{2} = 1$ for the unit area. The mass unit $\mu$ is set to $1$,~\textit{i.e.}, a cube of length $\ell^{3} = 1$ has a mass of $\mu = 1$. Finally, we set the unit of energy to $\varepsilon = k_{\mathrm{B}}T$, with $k_{\mathrm{B}}$ Boltzmann's constant and $T$ the temperature. This implies that the unit of time is $\tau \equiv \sqrt{\mu/\varepsilon} = 1$.

The dynamics of each triangle are specified by a strongly damped Langevin equation acting on the center of mass, to which the other beads spanning the triangle are rigidly connected. If the $i$-th triangle's position is denoted by $\boldsymbol{r}_{i}$, then the translational equation of motion for the center of mass becomes
\begin{align}
  \nonumber M_{i} \frac{\partial^{2}}{\partial t^{2}} \boldsymbol{r}_{i} &= - \underline{\boldsymbol{\Gamma}}_{t} \frac{\partial}{\partial t} \boldsymbol{r}_{i} + f \boldsymbol{\hat{u}}_{i} \\
  \label{eq:EOMt} &\quad - \sum_{j \ne i} \boldsymbol{\nabla} V(r_{ij},\boldsymbol{Q}_{i},\boldsymbol{Q}_{j}) + \boldsymbol{\xi}_{i,t}(t),
\end{align}
where $M_{i}$ is the triangle's mass; $\underline{\boldsymbol{\Gamma}}_{t}$ is the translational diffusion coefficient matrix, which accounts for shape anisotropy; $\boldsymbol{\hat{u}}_{i}$ denotes the triangle's orientational unit vector, which co-moves and co-rotates with the triangle; $f$ is the self-propulsion force; $\boldsymbol{\nabla}$ denotes the gradient; $V$ is a pair potential depending on the triangle separation $r_{ij} \equiv \vert \boldsymbol{r}_{i} - \boldsymbol{r}_{j} \vert$, and the orientation of both triangles, specified here by quaternions $\boldsymbol{Q}_{i}$ and $\boldsymbol{Q}_{j}$~\footnote{Note that our equation of motion is 3D and we therefore must employ quaternions to specify 3D rotation.}; and $\boldsymbol{\xi}_{i,t}(t)$ is the random translational thermal noise, which satisfies $\langle \boldsymbol{\xi}_{i,t}(t) \rangle = \boldsymbol{0}$ and $\langle \boldsymbol{\xi}_{i,t}(t) \otimes \boldsymbol{\xi}_{j,t}(t') \rangle = 6 k_{\mathrm{B}}T \underline{\Gamma}_{t} \delta_{ij} \delta(t - t')$, with $\otimes$ the dyadic product, $\delta_{ij}$ the Kronecker delta, and $\delta$ the one-dimensional (1D) delta distribution; $\langle \cdots \rangle$ indicates time averaging. 

The equation of motion for quaternions is similar to Eq.~\eqref{eq:EOMt}, but is not reproduced here, due to its length --- we refer the interested reader to Ref.~\cite{martys99} for the full details. Finally, the motion of the triangles is fixed to a 2D (planar) geometry with periodic boundary conditions. The triangles are aligned with the $xy$-plane and their centers can only translate in this plane; rotation is only permitted around the $z$-axis. This choice implies that the orientation of our triangles can also be described using a simple angle $\theta$ with respect to the $y$-axis, $\theta = \cos^{-1}\left( \boldsymbol{\hat{u}} \cdot \boldsymbol{\hat{y}} \right)$, and associated equation of motion. However, as we employed 3D algorithms together with constraints, we provided the full details here, rather than the reduced model the constraints lead to.

The reason for applying a damped, but not overdamped dynamics in our simulations is that coupling to say a lattice-Boltzmann fluid requires damped dynamics,~\textit{e.g.}, see Refs.~\cite{ahlrichs99,lobaskin04,roehm14}. Our present choice thus facilitates future comparison to systems with hydrodynamic interactions at a later stage~\cite{fischer15,degraaf15b,degraaf16a,degraaf16b}, but this goes beyond the scope of the current paper. However, the damping factor is strong, leading to a close to Brownian dynamics simulation, which is appropriate for the colloidal length scale. We make use of the software package \textsf{ESPResSo}~\cite{limbach06a,arnold13a} throughout to perform our simulations.

\subsection{\label{sub:params}Setup and Parameters}

\begin{figure}[!htb]
\begin{center}
\includegraphics[scale=1.0]{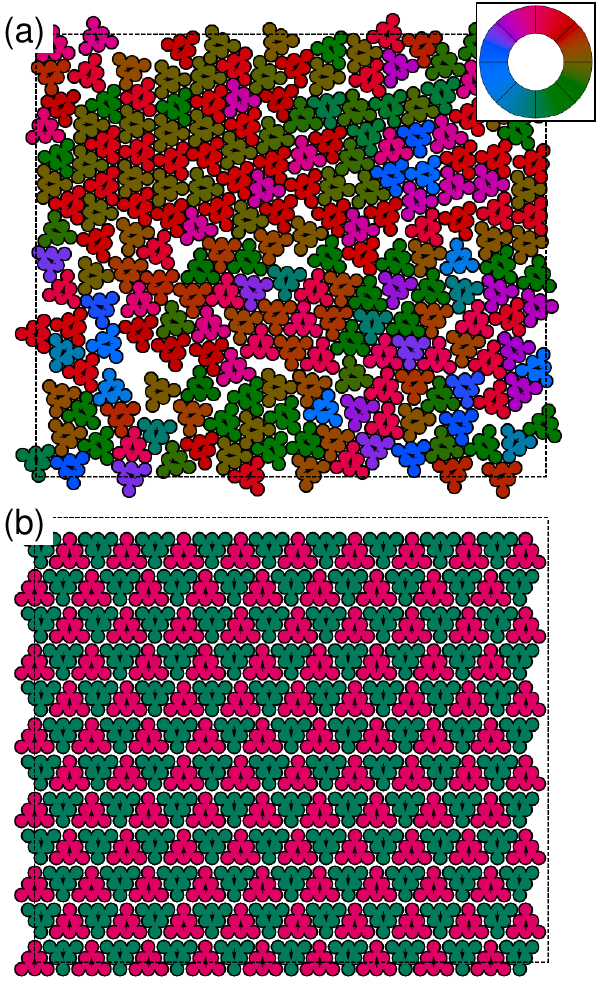}
\end{center}
\caption{\label{fig:setup}The two initial configurations used in our system for a triangle density of $\rho = 0.13$: (a) a random configuration and (b) a crystalline setup. The dashed line shows the outline of our simulation box. The color wheel in the inset to (a) serves as a legend to our orientation-based color scheme.}
\end{figure}

To study the MIPS of rough triangles, we considered rectangular simulation boxes containing $N_{T} = 216$ triangular particles, see Fig.~\ref{fig:setup}. This number may seem low, but because of our construction, we simulate $7N_{T} = 1512$ MD beads for $N=3$, and $16N_{T} = 3456$ for a discretization of $N=6$.

We used a fixed aspect ratio for the box dimensions that is commensurate with the closest packing of these triangles,~\textit{i.e.}, the case where they just touch with their interaction cut-off of $2^{1/6} \sigma$ ($\sigma$ is used here, because the cut-off value depends on the choice of $N$). Figure~\ref{fig:setup}b shows a close-to-dense crystalline assembly of triangles with an area density of $\rho = 0.13$ (number of particles per unit area $\ell^{2}$). We considered densities in the range of $\rho = 10^{-3}$ to $\rho = 0.14$, where the latter is approximately the limiting density --- the exact number depends slightly on the specific value of the roughness parameter $N$.

The physical properties of the triangles and the system were further specified by choosing the Langevin parameters. We chose translational friction coefficients for the triangle of $\Gamma_{t} = 10$ in both the $x$- and $y$-direction. We do not need to specify the value for the $z$-direction, due to the confinement and we assume there to be no cross-diffusional terms in $\underline{\boldsymbol{\Gamma}}_{t}$. The unusually high value of the friction coefficient was chosen to achieve a relatively low diffusivity for $\varepsilon = k_{\mathrm{B}}T$. This value allows us to obtain greater values of the P{\'e}clet number, for relatively low values of the propulsive force $f$, such that our algorithm remains numerically stable. 

There is a measure of arbitrariness in obtaining a rotational friction coefficient around the $z$-axis for the triangle. Fluctuation-dissipation states that hydrodynamic friction and diffusion are related via $\Gamma = k_{\mathrm{B}}T/D$, with $\Gamma$ the friction and $D$ the diffusivity. So a natural way to approach the problem would be to determine the hydrodynamic friction of a triangle, both translationally and rotationally, and use the ratio to determine $\Gamma_{r}$ from our choice of $\Gamma_{t}$. Unfortunately, the Stokes' paradox means that there is no solution for 2D hydrodynamic flow around a disk (or a triangle). Therefore, one must perform such a calculation in 3D instead, with a \textit{finite} triangular prism. The height of this prism introduces the arbitrariness. Here, we chose to use the ratio $\Gamma_{r}/\Gamma_{t} \approx 10$, which is obtained for a 3D cylinder segment with diameter $2R = 2L/\sqrt{3}$ and height $H = 2^{1/6}$ by making use of the expressions of Ref.~\cite{ortega03}. We use a cylinder segment instead of the triangular prism, because the triangular prism problem is much more difficult to solve and the level of arbitrariness in choosing $H$ does not justify further complicating the problem. This line of argument leads to a choice of $\Gamma_{r} = 100$ for rotation around the $z$-axis. The ratio $\Gamma_{r}/\Gamma_{t}$ is more representative of what one may encounter in an experimentally synthesized colloidal triangle, which has a finite width (about half the diameter), than say $\Gamma_{r}/\Gamma_{t} = 1$ would be. However, we should stress that other choices can be made, depending on the situation of interest.

As mentioned earlier, we use a thermal energy of $k_{\mathrm{B}}T = \varepsilon = 1$ for the Langevin thermostat. In order to eliminate the arbitrary choice of the force $f$ (the dimensionful quantity can be recovered by multiplying with $\mu \ell \tau^{-2}$), we use the P{\'e}clet number $Pe$ throughout. Here, the relation between $f$ and $Pe$ is as follows 
\begin{align}
\label{eq:Pe} Pe &= \frac{ L u }{ D_{t}} = \frac{L f}{k_{\mathrm{B}}T} = 2^{1/6}(2 + \sqrt{3})f,
\end{align}
where we chose the edge length $L$ to represent the relevant length in the definition of the $Pe$ number; $u$ is the self-propulsion speed of the particle, which is simply given by $u = f/\Gamma_{t}$; and $D_{t} = k_{\mathrm{B}}T/\Gamma_{t}$ per definition. This leads to the second and third expression in Eq.~\eqref{eq:Pe} and we verified this to be accurate to within 5\% in our numerical simulations. We used a maximum of $\vert f \vert = 50$, corresponding to $Pe \approx 210$. Our parameter choices allowed us to use a time step of $\Delta t = 0.01$, without running into algorithmic instabilities.

For all systems with density $\rho < 0.1$, we initialized the triangles randomly and without overlap, see Fig.~\ref{fig:setup}a. For systems above this density we considered a crystalline setup, which is visualized in Fig.~\ref{fig:setup}b. This is because for the passive (\textit{not} self-propelled) particles it is difficult to achieve an initial non-overlapping configuration when $\rho \ge 0.1$. Each simulation consisted of an equilibration and production phase. The system was equilibrated for $10^{6}$ time steps of $\Delta t = 0.01$. For the production runs we used $5\cdot10^{6}$ time steps of the same size, leading to a total production time of $5\cdot10^{4}$ in MD units. During production we sampled the measureable quantities and took 500 snapshots of the simulation (in some cases 5000). The quantities that we obtained are discussed in the following.

\subsection{\label{sub:charac}System Characterization}

To characterize the physics of the system, we considered several methods. First, we scrutinized the system visually, making use of orientation-based color coding (see the inset to Fig.~\ref{fig:setup}a) to obtain an overview of the major features. These features are subsequently characterized by means of the radial distribution (pair correlation) function between the centers of the triangles, as defined by
\begin{align}
\label{eq:rdf} g(r) &= \frac{1}{2 \pi r dr} \frac{1}{2 \rho N_{T}} \left\langle \sum_{i = 1}^{N_{T}} \sum_{j \ne i}  \delta \left( r - r_{ij} \right) \right\rangle,
\end{align} 
where $r$ is radial distance, $dr$ is the bin width. The radial distribution function can be used to determine whether the system is crystalline or liquid-like. However, due to the surface roughness, there is additional local structure on top of the typical minima and maxima in $g(r)$ that one may expect for a crystalline phase; we therefore do not present these results here.

By determining the first minimum in $g(r)$ the average nearest-neighbor distance $r_{\mathrm{nn}}$ can be extracted. This in turn was used to compute the bond-orientational order parameter
\begin{align}
\label{eq:bondord} \Psi_{n} = \frac{1}{N_{T}} \left\langle \sum_{i = 1}^{N_{T}} \bigg\vert \frac{1}{N_{\mathrm{b}}} \sum_{j=1}^{N_{\mathrm{b}}}  \exp( \imath n \theta_{j} ) \bigg\vert \right\rangle,
\end{align}
where $N_{\mathrm{b}}$ is the number of nearest neighbors of particle $i$ ($\vert r_{ij} \vert < r_{\mathrm{nn}}$), $\imath$ the imaginary unit, $\theta_{j}$ is the angle between an arbitrary fixed direction $\boldsymbol{\hat{n}}$ and the bond vector between particle $i$ and $j$, and $\vert \cdots \vert$ indicates the absolute of a complex number. Important for this definition is that the angle is always measured in a clockwise (or counterclockwise) direction starting from the arbitrary direction $\boldsymbol{\hat{n}}$, which we choose to be the unit vector pointing along the $y$-axis ($\boldsymbol{\hat{y}}$) here. We restrict ourselves to $\Psi_{3}$, as it gives the most insight into the system. The signature of clustering is strongly pronounced using $\Psi_{3}$, due to the triangular symmetry of the particles. 

Finally, we compute the cluster distribution in the system. Determining clusters in the system is difficult, because of the triangular shape, the surface roughness, and the ``softness'' of the WCA potential. Both lead to variation in the separation between triangles that are clustered, but similar values of the separation are found for misaligned triangles. Using simply the first minimum in $g(r)$, as for our $\Psi_{3}$ analysis, proved an inadequate criterion to establish clustering. Following careful study of many configurations, we obtained a center-to-center distance of $r_{c} = 2.41$ (for $N=3$; $L = 4.19$) such that the number of instances where a triangle is erroneously identified as (not) being a part of a cluster is relatively small. 

To determine the state diagram using our algorithm, the clusters were subsequently binned by size and the number of clusters per bin was averaged over the span of the simulation. These averaged values were reweighted by the total number of clusters of that size that can be present in the system. That is, there can be $N_{T}$ unclustered triangles, $N_{T}/2$ dimers, ..., and 1 system spanning cluster. The binned, averaged, and reweighted values are referred to as cluster numbers $C_{i}$, with $i = 1$ for monomers, $i = 2$ for dimers, $\dots$, and $i = N_{T}$ for system spanning clusters.

We further used our cluster criterion to determine average life span $t_{\mathrm{av}}$ of small clusters and establish their relative frequency $f$. Finally, for system-spanning clusters, we determined the probability $F$ of finding a cluster that persisted for a time $t_{p}$. We did this by binning and weighting the times for which more than half of the system was covered by a single cluster.

\section{\label{sec:result}Results}

In this section we discuss the main results of our simulations. We start with the visualization of the state diagram. This is followed by a more rigorous analysis based on our cluster criterion. Next we examine the shape and size of the small clusters observed at low triangle densities as well as the breakup of larger clusters comprised of BD triangles. Finally, we consider the effect of surface roughness on our simulation results.

\begin{figure*}[!htb]
\begin{center}
\includegraphics[scale=1.0]{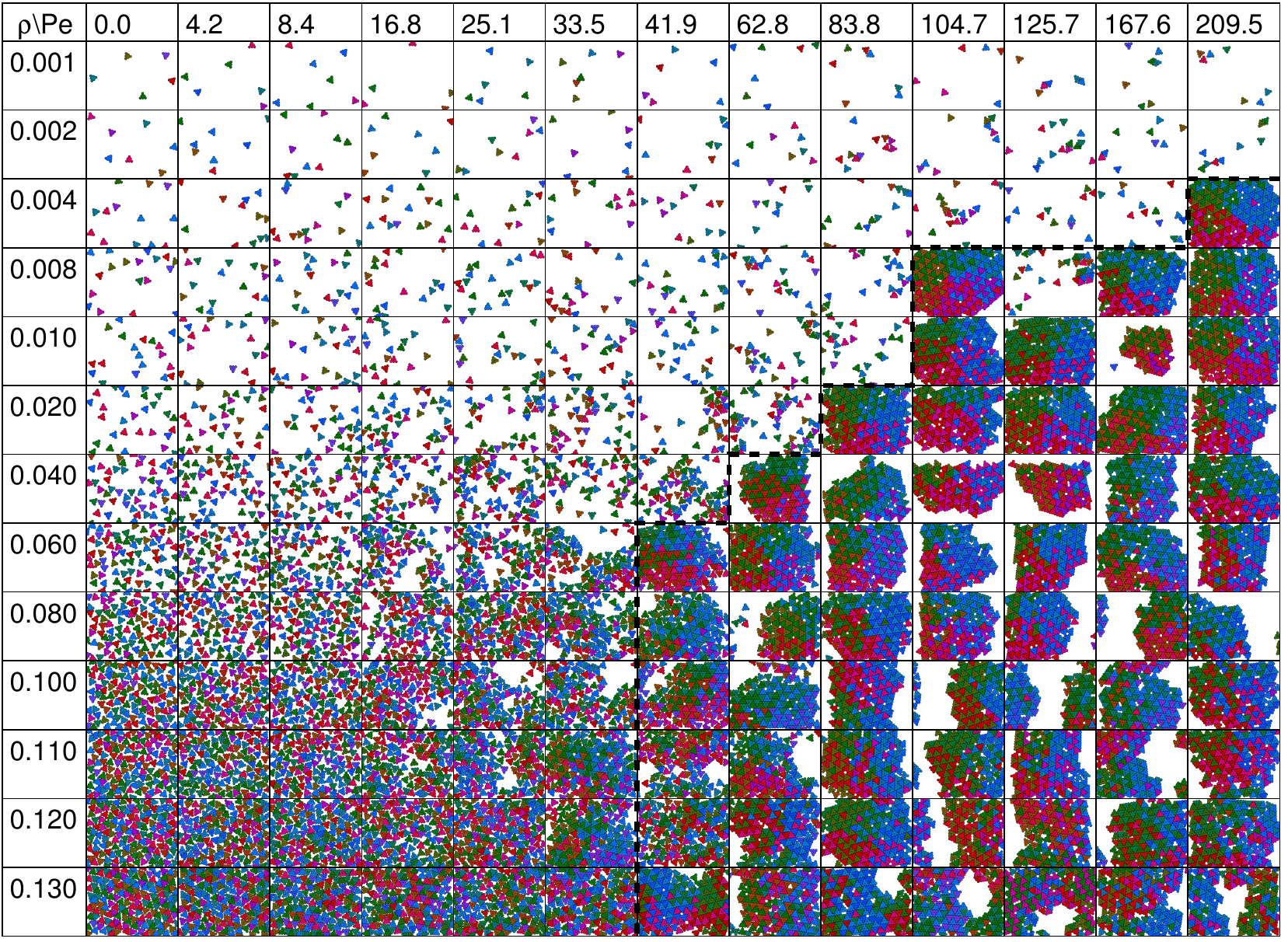}
\end{center}
\caption{\label{fig:forw}Representative snapshots of the clusters formed by AD triangles with $N=3$ as a function of the density $\rho$ (rows) and the P{\'e}clet ($Pe$) number (columns). Only a part of the simulation box is show in order to ensure that the size of the triangles is the same in all panels. The coloring shows the triangle orientation, see the inset to Fig.~\ref{fig:setup}a. The thick dashed line gives a rough indication of the region in which motility induced phase separation (MIPS) occurs.}
\end{figure*}

\begin{figure*}[!htb]
\begin{center}
\includegraphics[scale=1.0]{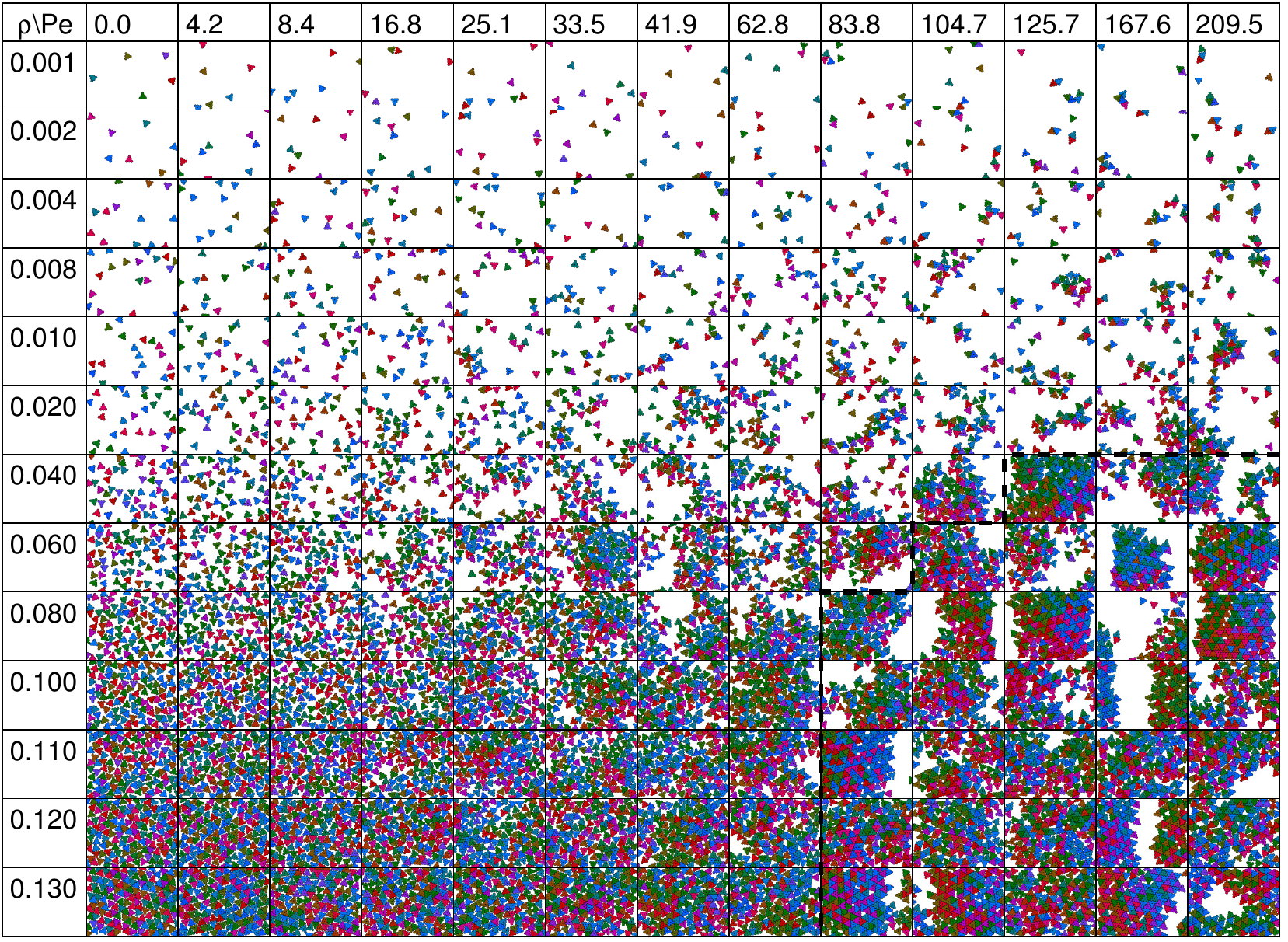}
\end{center}
\caption{\label{fig:back}Representative snapshots of the clusters formed by BD triangles with $N=3$ as a function of $\rho$ (rows) and P{\'e}clet ($Pe$) number. The notation is otherwise the same as in Fig.~\ref{fig:forw}.}
\end{figure*}

Figures~\ref{fig:forw} and~\ref{fig:back} show the representative snapshots of the state the system is in for AD and BD triangles, respectively, taken at the end of our simulations. Here, we processed the data in such a way that the field of view is the same size for each considered density,~\textit{i.e.}, the triangles are the same size. The window is centered on the region of the simulation box that contains the most triangular particles. There is a clear transition from systems without MIPS to a system with system-spanning clusters in both cases, with the clusters occurring for high $Pe$ and $\rho$. A rough divide based on our visual inspection of the systems is provided in Figs.~\ref{fig:forw} and~\ref{fig:back}. This shows that there is a larger region in the state diagram over which AD triangles form system spanning clusters than BD triangles.

The major difference between AD and BD triangles that can be observed from Figs.~\ref{fig:forw} and~\ref{fig:back} is that the clusters of AD triangles appear to have a ``smooth'' interface between the dense and dilute phase after MIPS. The BD clusters have an overall ``rougher'' interface and the clusters appear less densely packed. In fact, analysis of our data shows that the BD clusters break up relatively often compared to the AD ones, as we will come back to. This is why in Fig.~\ref{fig:back} there are more snapshots showing unclustered phases for high $Pe$ and $\rho$. Careful inspection of the clusters reveals that AD clusters have three to four distinct regions of triangle orientation, which on average point inward towards the center of the cluster. Within each region there are two orientations that differ by $60^{\circ}$, which are stacked alternatingly. The BD triangles show some of these features, but they are far less pronounced. 

Finally, we should note that for $Pe = 0$ and our highest density the system is not crystalline. This is not unreasonable, as the Mermin-Wagner theorem~\cite{mermin66} forbids spontaneous breaking of continuous symmetry in a finite-temperature 2D system with only short-ranged (and smooth) interactions. However, our systems have periodic boundary conditions, which would alleviate this constraint. To truly observe this effect we would need to simulate prohibitively large system sizes. For active systems the Mermin-Wagner theorem does not apply and MIPS can occur. We have examined systems with smaller numbers of particles and the features shown in Figs.~\ref{fig:forw} and~\ref{fig:back} are largely preserved. This indicates that simulating 216 triangles is sufficient to obtain a qualitative understanding of the behavior of the system.

\begin{figure}[!htb]
\begin{center}
\includegraphics[scale=1.0]{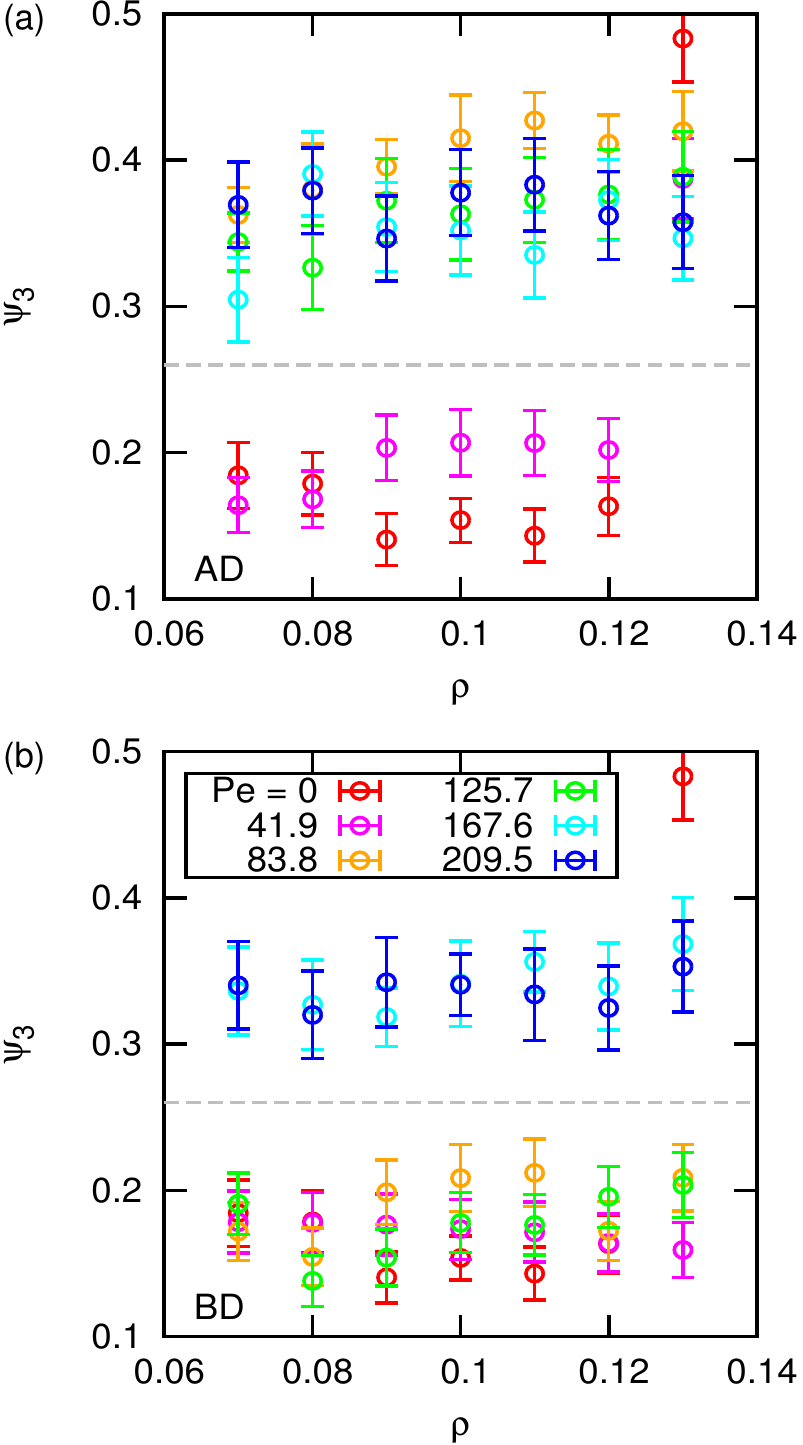}
\end{center}
\caption{\label{fig:psi}The value of the three-fold bond-orientational order parameter $\Psi_{3}$ in the clustering region as a function of the density $\rho$ for several values of $Pe$ for AD triangles (a) and BD triangles (b). The grey dashed horizontal line is a guide to the eye to distinguish non-clustered systems from those that have clustered. In both panels the value of $\Psi_{3}$ of the passive system ($Pe = 0$) for $\rho = 0.13$ is an outlier.}
\end{figure}

For the systems shown in Figs.~\ref{fig:forw} and~\ref{fig:back}, we considered the three-fold bond-orientational order parameter $\Psi_{3}$ to differentiate between clustered and non-clustered systems. It proved difficult to obtain satisfactory statistics for the BD triangles due to the constant breaking and reforming of the clusters, as well as for the AD clusters in the close-to MIPS region. However, the trends observed Figs.~\ref{fig:forw} and~\ref{fig:back} were recovered in our analysis, see Fig.~\ref{fig:psi}. Above a threshold $Pe$ the system clusters. For the BD triangles this clustering is shifted towards higher values of the $Pe$. Finally, it should be noted that the value of $\Psi_{3}$ of the passive system ($Pe = 0$) at high densities is an outlier in each graph. This is due to the passive system having a ``more uniform'' crystalline state than its active counterparts. That is, for the active systems there is an interface in the system for high $\rho$ and $Pe$, which reduces the overall value of $\Psi_{3}$, through the lower local $\Psi_{3}$ at the boundary. 

\begin{figure*}[!htb]
\begin{center}
\includegraphics[scale=1.0]{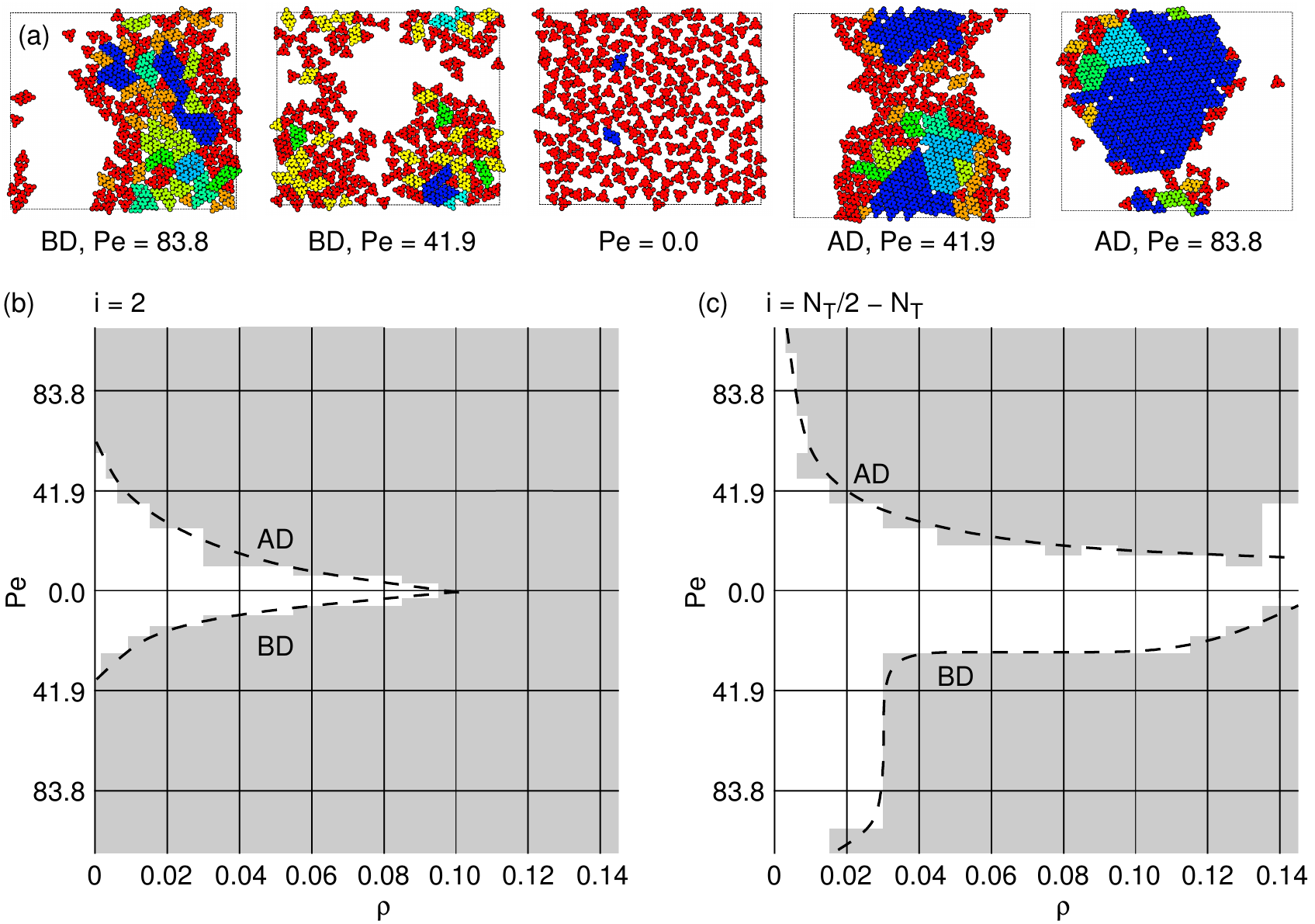}
\end{center}
\caption{\label{fig:cluster}The cluster criterion used to determine properties of the system. (a) Representative snapshot of a $N=3$ system with density $\rho = 0.08$. From left to right, the self-propulsion goes from BD to AD. The coloring is based on the cluster algorithm, red are single particles and blue represents the largest cluster in the system. From red to blue the size of the clusters increases, but not according to fixed increments. N.B. The magenta coloring can be difficult to distinguish from red. (b-c) Cluster diagrams in the ($\rho,Pe$)-representation. The top part corresponds to AD, the bottom part to BD, and $Pe = 0.0$ to passive triangles. The gray region shows where the value of the cluster number $C_{i} > 0.02$. The dashed lines serve as guides to the eye. (b) The region where dimers form ($i = 2$). (c) The region where the cluster becomes system spanning ($i$ binned from $N_{T}/2$ to $N_{T}$).}
\end{figure*}

Since our visualizations and the values of the bond-orientational order parameter $\Psi_{3}$ only give limited insight into our system, we also considered a cluster criterion, as described in Section~\ref{sub:charac}. Figure~\ref{fig:cluster}a shows the result of our cluster criterion for a number of snapshots of systems with varying $Pe$ at a density of $\rho = 0.08$. Close inspection of the snapshots shows that there are some apparent mismatches. For example, several pairs of triangles for $Pe = 0$ that have not been identified as a dimer. There is a small misalignment or larger gap between the two triangles in these cases, than there is for the two pairs that have been identified as dimers. This responsiveness to small changes is a consequence of priming our cluster criterion,~\textit{i.e.}, it can distinguish between a dense cluster and a ``loose assembly'' at $\rho = 0.13$. Comparison between the snapshots with $Pe \approx 42$ and $Pe \approx 83$ shows that the algorithm effectively separates the dense crystalline regions from those which are not crystalline.

Using the algorithm we were able to draw a \textit{tentative} state diagram for the MIPS of AD and BD triangles with a roughness parameter of $N=3$, see Fig.~\ref{fig:cluster}. Here we used a cut-off for the cluster number of $C_{i} > 0.02$ to determine whether there is significant formation of dimers and system-spanning clusters, respectively. Different values of the cut-off shift the boundaries of the diagram, but leave the trends preserved. A low value was necessary due to the BD-triangle clusters often breaking up, as we will come back to. Our cluster criterion clearly picks up on the asymmetry between the clustering of AD and BD triangles that was also apparent in Figs.~\ref{fig:forw}~and~\ref{fig:back}. Also note that our criterion shows that there is no system-spanning cluster for $Pe = 0$, which is to be expected, and thus gives a better insight into the system than $\Psi_{3}$. There is some statistical noise on the data at the edge of the cluster domain, which is due to the clusters breaking up and reforming quite frequently in this region --- breakups appear to be more frequent for higher $Pe$.

The above asymmetry between AD and BD triangles is similar to the one observed by Wensink~\textit{et al.}~\cite{wensink14}. However, Wensink~\textit{et al.} observe ``\textit{Polar $s^{+}$ SPPs form aligned large-scale swarms that move cooperatively along a spontaneously chosen common axis}'' and  ``\textit{antipolar $s^{−}$ SPPs tend to form droplets that nucleate slowly from an initially homogeneous suspension},'' where polar particles correspond to our AD and antipolar particles to our BD triangles, respectively, and SPP stands for self-propelled particle. This is the complete opposite of our finding, since our AD (polar) triangles form clusters more readily than the BD ones. The reason for the difference must be due the shape of our polar particles compared to those of Ref.~\cite{wensink14}, as we will explain next.

The polar SPPs of Ref.~\cite{wensink14} are much rounder and the angle of their ``tapered end'' is more acute than that of our triangles. We presume that this roundness aids the formation and compactification of clusters of antipolar particles, compared to our triangular swimmers. For polar swimmers, the acuteness of the tapering and the roundness of the particle, presumably lead to less stable small structures being formed, from which stable larger clusters can nucleate. Our AD triangles are tailored to facilitate the formation of hexagonal composites, as we will come back to, which may give the stability needed for the AD particles to cluster further. It could be argued that the roughness of our particles can also play an important role in the difference between our results and those of Wensink~\textit{et al.}. However, as we will see, smoothing of the triangular edges only further destabilizes the BD clusters, and the primary effect is thus the shape difference.

\begin{figure*}[!htb]
\begin{center}
\includegraphics[scale=1.0]{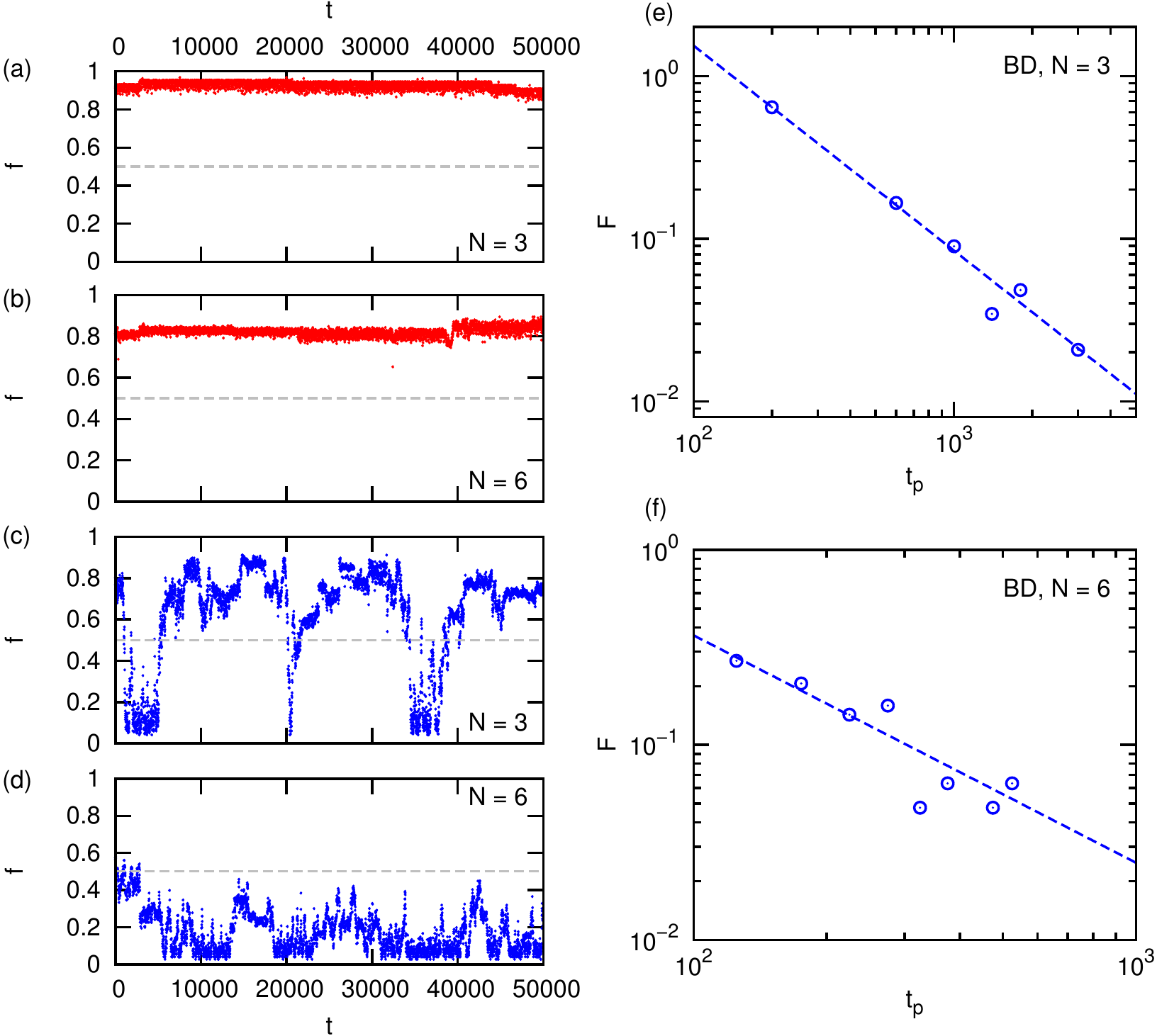}
\end{center}
\caption{\label{fig:break}The stability of system spanning clusters in a system with $\rho = 0.07$ and $Pe \approx 210$. (a-d) The fraction $f$ of the system occupied by the largest cluster as a function of time $t$ for both AD (a,b) and BD (c,d) triangles and different levels of roughness $N=3$ (a,c) and $N=6$ (b,d). Here, the largest cluster includes all the pieces for which our algorithm detects the grain boundaries, e.g., see Fig.~\ref{fig:cluster}a. (e) The fraction $F$ of system spanning clusters that persist for a time $t_{p}$ out of all system spanning clusters observed for BD triangles with $N=3$. The data (blue circles) is binned with bin size $400$ from $t_{p} = 200$ to $5000$ and plotted on a log scale. The blue dashed line gives a power-law fit to the data. (f) $F$ as a function of $t_{p}$ for BD triangles with $N=6$. Here, the data was again binned with bin sizes of $50$ from $t_{p} = 125$ to $1025$.}
\end{figure*}

Let us now turn our attention to the stability of the system-spanning AD and BD clusters. Figure~\ref{fig:break}a-d show the fraction $f$ of the system occupied by the largest cluster as a function of the simulation time $t$ for AD (a,b) and BD (c,d) triangles in a system with $\rho = 0.07$ and $Pe \approx 210$. For the $N = 3$ triangles, we observe that the AD triangles form a system-spanning cluster that is stable throughout the simulation time, whereas similar clusters of BD triangles periodically break up. We consider the results for $N=6$ shortly. We determined the length for which the system-spanning cluster persists $f > 0.5$, by performing 10 runs similar to the one in Fig.~\ref{fig:break}a,c. For AD triangles the persistence time $t_{p}$ is equal to the length of the simulation. Binning and weighting the results for BD triangles gives the fraction $F$ of observed clusters that persist for $t_{p}$. Most clusters persist around $t_{p} \approx 400$, with the probability of finding a cluster that persists longer decaying as a power law. In each case, the system-spanning cluster of BD triangles breaks up by rotating at increasing speed, before coming apart in its entirety. 

It should also be noted that the breakup of large clusters has been observed in systems consisting of active squares~\cite{prymidis16}, where these events where referred to as oscillations. Prymidis~\textit{et al.}~\cite{prymidis16} also observed that the breakup of clusters is preceded by the cluster rotating as a whole. For the active squares, a completely homogenous system was recovered after breakup, whereas our systems appear to have local density variations. However, it is difficult to say on the basis of our results to what extent this is due to the relatively low number of triangles simulated. The state diagram of Ref.~\cite{prymidis16} shows that such oscillatory events take place over a large band in density and activity. In addition, for the active squares, above a certain $\rho$ and $Pe$ critical line, a transition from the oscillatory state to a stable clustering state is achieved. The breaking up of clusters for relatively high $\rho$ and $Pe$ in the Fig.~\ref{fig:break}c,e, as well as for other combinations of $\rho$ and $Pe$ shown in Fig.~\ref{fig:back}, suggests that we have not yet entered this stable clustering regime, for any of the $\rho,Pe$ combinations that we have simulated, if it even exists for our particles.

\begin{figure}[!htb]
\begin{center}
\includegraphics[scale=1.0]{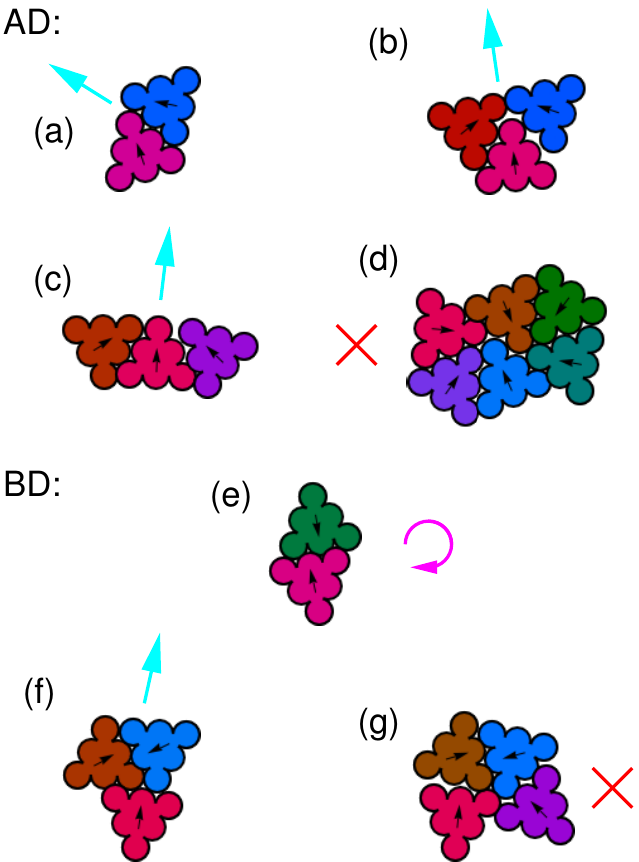}
\end{center}
\caption{\label{fig:small}Representative samples of the small clusters that are formed at low density ($\rho = 0.002$, $Pe \approx 210$); (a-d) formed by AD triangles and (e-g) formed by BD triangles. (a) A dimeric cluster of AD triangles, which moves in the direction of the resultant force vector, as indicated by the cyan arrow. (b-c) Two AD trimers, both of which move in the direction of the resultant force vector. (d) A hexamer comprised of 6 AD directed triangles, which exhibits only limited translation and rotation, as indicated using the red cross. (e) A dimer comprised of two BD triangles. This type of cluster rotates, as indicated by the magenta arrow. (f) A trimer of BD triangles, which on average moves in the direction of the resultant vector. (h) A quadrumer of base directed triangles, which shows only very limited movement.}
\end{figure}

For both AD and BD triangles small clusters form at low density, which provide insight into the system-spanning clusters that form at higher density. Small oligomers consisting of up to 6 triangles are visualized in Fig.~\ref{fig:small}. AD triangles form dimers, where the triangles push into each other sideways and the composite moves forward along the resultant force vector, as shown in Fig.~\ref{fig:small}a. AD triangles also form two types of trimer, see Fig.~\ref{fig:small}c,d, both travelling in the direction of the resultant force vector. Finally, Fig.~\ref{fig:small}d shows an AD hexamer with all triangle apexes pointing inward. These type of clusters only move and rotate slowly --- the roughness of the triangles prevents exact cancellation of the self-propulsive forces. 

BD triangles also form dimers, for which the bases touch, leading to the pair spinning rapidly, without translating, see Fig.~\ref{fig:small}e. BD dimers can be joined by another triangle (Fig.~\ref{fig:small}g) or even two triangles (Fig.~\ref{fig:small}h) to form composites that again translate slowly or are more or less immobilized, respectively.  

\begin{figure}[!htb]
\begin{center}
\includegraphics[scale=1.0]{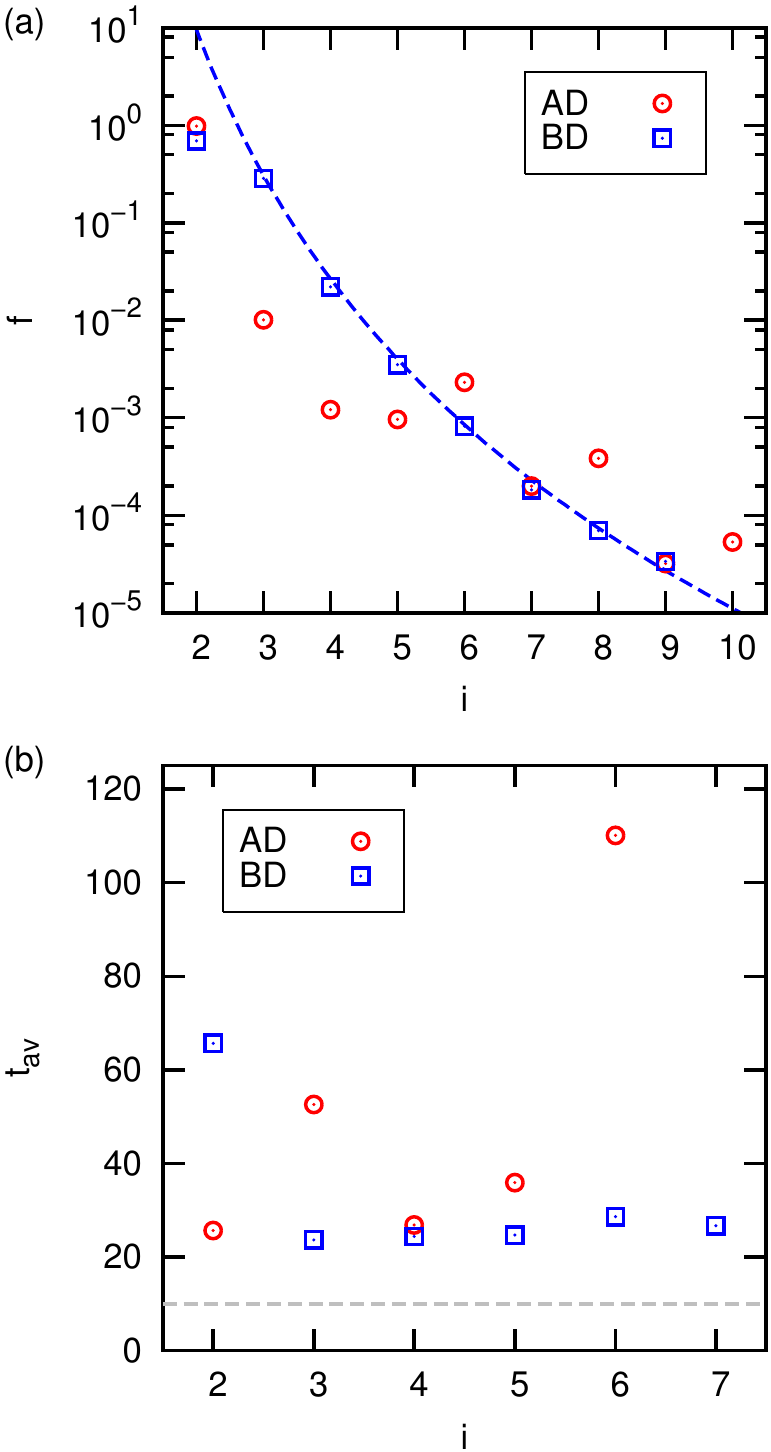}
\end{center}
\caption{\label{fig:times}Properties of the small clusters at low density ($\rho = 0.002$, $Pe \approx 210$). (a) The fraction $f$ of observed clusters with size $i$. The symbols indicate the cluster size, the dashed curve is a guide to the eye for the power-law decay of the BD-triangle cluster size. (b) The average cluster persistence time $t_{\mathrm{av}}$ as a function of the cluster size $i$. The dashed gray horizontal line shows the time cut-off for cluster acquisition.}
\end{figure}

To quantify the properties of these small clusters, we analyzed their frequency $f$ and average life span $t_{\mathrm{av}}$ as a function of the cluster size $i$, see Fig.~\ref{fig:times}. Figure~\ref{fig:times}a, shows the relative frequency of a cluster of size $i$ with respect to the total number of clusters observed. For both AD and BD triangles, the dimers are the most frequent, but for BD triangles trimers are also observed relatively often. For BD triangles, $f$ decreases as a power law with $i$, as indicated by the dashed blue line, for $i \ge 3$, while for AD triangles, the trend is less obvious. Finally, for AD triangles larger oligomers, especially hexamers, are more frequent. 

We examined the life span $t_{\mathrm{av}}$ of these clusters, as shown in Fig.~\ref{fig:times}b. In most cases, the time $t_{p}$ a small cluster persisted decayed exponentially. To determine $t_{\mathrm{av}}$ a probability density functional for $t_{p}$ was established and the expectation value $t_{\mathrm{av}}$ was computed. Our cluster algorithm used snapshots spaced $\Delta t = 10$ apart, which means that $t_{p} = 10$ is the minimum persistence time, as indicated in Fig.~\ref{fig:times}b by the dashed gray line. It is clear that for most clusters $t_{\mathrm{av}} \approx 25$, but there are a few outliers. For AD triangles, trimers and hexamers are more stable than other cluster sizes, with the hexamers having an average life span that is $\approx 4$ times the average. This lends credibility to our hypothesis that AD hexamers are the stable nucleus around which larger clusters can form. Only the dimer is relatively stable for BD triangles.  

\begin{figure*}[!htb]
\begin{center}
\includegraphics[scale=1.0]{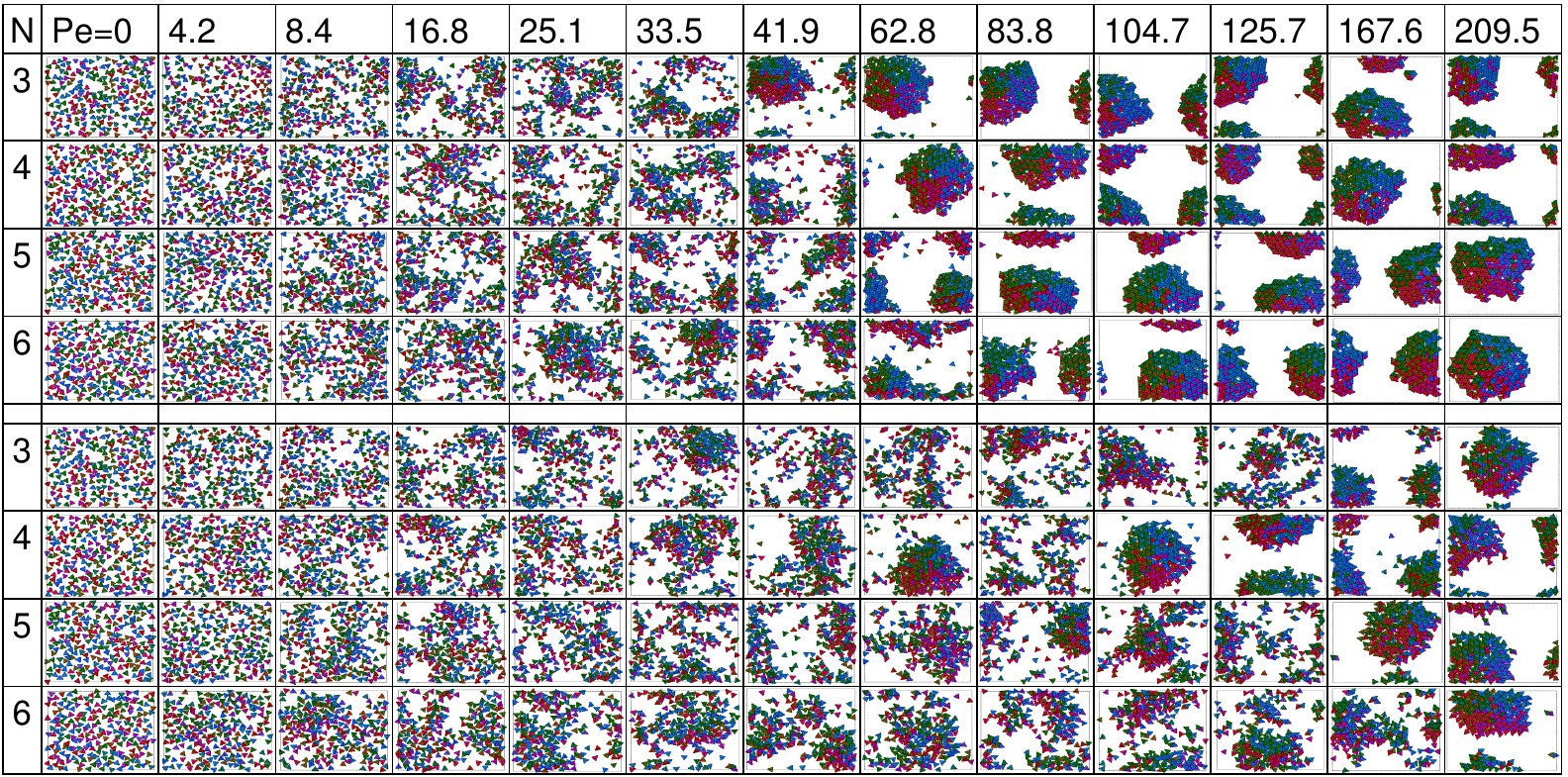}
\end{center}
\caption{\label{fig:disc}Representative snapshots of the clusters formed by AD triangles with $\rho =0.06$ as a function of roughness parameter $N$ (rows) and $Pe$ (columns). The top four columns give the results for AD triangles, while the bottom four rows the result for BD triangles. The entire simulation box is shown in each panel and the coloring corresponds to the triangle orientation, see the inset to Fig.~\ref{fig:setup}a.}
\end{figure*}

Finally, we consider the effect of surface roughness on our state diagram. We examine one intermediate density specifically, namely $\rho = 0.06$, in Fig.~\ref{fig:disc}, which shows representative snapshots of the system. It is clear that for AD triangles, changing the edge roughness does not strongly influence the result of our simulations. However, for BD triangles, we found that clusters formed less frequently with decreased roughness. Only a few of the $N>3$ snapshots show system spanning clusters, and this trend is representative of all our simulations for these systems. The reduced clustering is presumably due to the smaller roughness negatively impacting the stability of the clusters as a whole. 

This result is further supported by our simulations for the system with $\rho = 0.07$ and $Pe \approx 210$, see Fig.~\ref{fig:break}. For AD clusters, increasing $N$ from $3$ to $6$ does not negatively impact the stability of the system-spanning cluster, while for BD clusters, system spanning clusters are virtually absent. That is, for $N=3$ BD triangles form a system spanning cluster $\approx 0.72$ of the total run time, whereas for $N=6$ this number has gone down to $\approx 0.06$. In addition, as can be seen by comparing Fig.~\ref{fig:break}e,f, the total time that clusters persist $t_{p}$ decreases much more rapidly for $N = 6$. 

A possible explanation of the reduced stability of the system-spanning BD clusters is as follows. At sufficiently high density, BD triangles can collide into composite objects and do not scatter off as easily as AD triangles, because the contact with the base of the triangle necessitates a larger rotation before the triangle can move away. This facilitates aggregation, but leads to rather rough clusters. Within such clusters, the only elements that prevent breakup are BD triangles that form dimers with the bases touching. Such dimers are quite stable, as they can only break up due to lateral (shear) forces, which are countered by the friction that roughness induces between the two touching faces. When the roughness is reduced, the friction force is lowered and the stabilizing dimers can more readily break apart. This then decreases the ability of BD to form clusters, as well as reduces the stability of those clusters that do form. 

Based on extrapolation our data we \textit{speculate} that in the flat-edged limit only clusters of AD triangles are stable. This need not be incommensurate with the observation of clustering for spheres and disks at sufficiently high $Pe$,~\textit{e.g.}, see Ref.~\cite{Stenhammar13}, as there is a significant difference in shape. In systems of rods, for instance, the only phase that has local crystallinity is the jammed phase~\cite{wensink12}, although the swarming phase could be considered a clustering phase as well, albeit one which exhibits directed motion and global orientational alignment. However, the observation of clustering for the system of self-propelled squares by Prymidis~\textit{et al.}~\cite{prymidis16}, suggests that this extrapolation may be too far reaching. Further investigation in the flat-edge limit is required to resolve this hypothesis.

\section{\label{sec:disc}Feasibility}

Before coming to the conclusions, it is important to comment on the experimental feasibility of realizing (triangular) particles with surface roughness. Colloidal particles with significant surface roughness, which are even comprised of spherical lobes as our triangles are, have recently been synthesized~\cite{kraft10,kraft13}. It is therefore conceivable that self-propelled particles with significant surface roughness can be prepared by capping one of the lobes with platinum and suspending these particles in hydrogen-peroxide solution. Furthermore, as already observed in Ref.~\cite{wensink14}, triangular particles can be considered models for self-propelled microorganisms such as sperm and \textit{Chlamydomonas} algae. 

In the current simulations, hydrodynamic effects were ignored. This is often considered a good approximation to study the behavior of self-propelled particles, as the major features, such as clustering, can be captured without the complication of simulating the fluid flow. However, it should be noted that near-field hydrodynamic and lubrication effects are always present in systems of colloidal particles suspended in fluid. These could modify the extent to which surface roughness is experienced between the particles. While the authors are not aware of any systematic experimental investigation of this for a system of self-propelled particles, lessons can be drawn from externally driven colloids. Specifically, recent experiments show that shear thickening in driven systems of suspended colloids is dominated by contact forces~\cite{lin15}. Transferring this finding to swimmers with surface roughness, the modification of the particle-particle contact through surface roughness could have a dominant effect over the hydrodynamic interactions.

Finally, we mention the effect of roughness in dry active matter. Typical examples of such systems are vibrated rods that belong in the granular regime of matter~\cite{deseigne10,kumar14}. In such systems, direct contact is made between the particles and the effects of roughness can be studied without the added complication of a suspending fluid. However, the specifics of the athermal translational and rotational diffusion will be different from those considered in our work, as the fluctuation-dissipation does not hold.

\section{\label{sec:conc}Conclusion and Outlook}

Summarizing, in this paper we have studied the 2D motility induced phase separation (MIPS) of active equilateral triangles, which self-propel either towards their apex (AD) or their base (BD). These particles have a polarity due to their shape and the direction of self-propulsion, similar to the polar swimmers considered by Wensink~\textit{et al.}~\cite{wensink14}. Our triangles in addition have a surface roughness. This roughness is caused by the discretization of the triangle edges using spheres and it provides us with a second variational parameter. 

We constructed state diagrams for both AD and BD triangles by varying the density $\rho$ and P{\'e}clet ($Pe$) number. We find MIPS for both systems, and characterized it using three different methods. In all cases we found an asymmetry between the clustering behavior of AD and BD triangles. AD triangles tended to phase separate considerably lower Pe numbers and densities than their BD counterparts. Surprisingly, our asymmetry is in the opposite direction of that observed by Wensink~\textit{et al.}~\cite{wensink14}. Namely, AD triangles clustered more readily than their BD counterparts, whereas the polar (AD) swimmers of Ref.~\cite{wensink14} did not cluster, when their antipolar (BD) swimmers did. The reason for this difference is due to the difference in shape between our and their particles and illustrates the sensitivity of the MIPS behavior to this parameter.

Both AD and BD triangles formed a variety small clusters at low $\rho$. The AD hexagonal clusters appeared to be the stabilizing element in inducing further growth of clusters. For BD triangles the clusters are stabilized by contact with the base of the triangle, as this necessitates a larger rotation before the triangle can move away from a cluster than for an AD triangle. Such a larger rotation also means that the triangles become a part of a cluster when they are more misaligned than would be possible for their AD counterparts. This prevents cluster growth at intermediate densities, as the newly incorporated elements are easily knocked off, due to their loose connection with the cluster. 

The system-spanning clusters formed by BD triangles had a rougher interfacial structure, and BD clusters broke up quite frequently. This limited stability and rougher interface can be explained by the clusters having a coarser internal structure, due to the relative ease with which BD triangles are incorporated into clusters. We speculate that there are analogies between the oscillating phases observed by Prymidis~\textit{et al.}~\cite{prymidis16} in a system of self-propelled squares and our frequent breaking up of the BD triangular clusters. In both cases, breakup of the cluster is preceded by the cluster rotating as a whole.

Finally, we found that the stability of the BD-triangle clusters is reduced by decreasing the triangle surface roughness. This may be due to the cluster's dimeric ``bonds'' being more susceptible to lateral forces, for a smoother triangle surface. It is therefore possible that in the flat-edged limit, only the AD system can form large stable clusters. However, this is a controversial statement, as clustering is observed in many self-propelled systems at sufficiently high $Pe$. Careful study of a flat-edge model is required to (dis)prove this claim, which unfortunately goes beyond the scope of this paper.

Future investigations will focus on the study of systems of triangles which interact via hydrodynamic flow fields. The description used in our work, in which the triangles are comprised of spheres, lends itself well to achieve hydrodynamic coupling with an lattice-Boltzmann fluid. In addition, we will further investigate the effect of smoothing of the triangles by exploring the flat-edge limit, in which the large particle limit is more readily achieved.

\section*{\label{sec:ack}Acknowledgements}

JdG gratefully acknowledges financial support by a Marie Sk{\l}odowska-Curie Intra European Fellowship (G.A. No. 654916) within Horizon 2020. JdG and CH thank the DFG for funding through the SPP 1726 ``Microswimmers --- From Single Particle Motion to Collective Behavior''. We are also grateful to F. Schultz and F. Weik for useful discussions.



\begin{thebibliography}{10}

\bibitem{ramaswamy10a}
S.~Ramaswamy,
\newblock Ann. Rev. Cond. Mat. Phys. {\bf 1}, 323 (2010).

\bibitem{marchetti13a}
M.~Marchetti et~al.,
\newblock Rev. Mod. Phys. {\bf 85}, 1143 (2013).

\bibitem{paxton04a}
W.~F. Paxton et~al.,
\newblock J. Am. Chem. Soc. {\bf 126}, 13424 (2004).

\bibitem{wang06a}
Y.~Wang et~al.,
\newblock Langmuir {\bf 22}, 10451 (2006).

\bibitem{howse07a}
J.~R. Howse et~al.,
\newblock Phys. Rev. Lett. {\bf 99}, 048102 (2007).

\bibitem{valadares10a}
L.~F. Valadares et~al.,
\newblock Small {\bf 6}, 565 (2010).

\bibitem{ebbens12a}
S.~Ebbens, M.-H. Tu, J.~R. Howse, and R.~Golestanian,
\newblock Phys. Rev. E {\bf 85}, 020401 (2012).

\bibitem{lee14a}
T.-C. Lee et~al.,
\newblock Nano Lett. {\bf 14}, 2407 (2014).

\bibitem{brown14a}
A.~Brown and W.~Poon,
\newblock Soft Matter {\bf 10}, 4016 (2014).

\bibitem{ebbens14a}
S.~Ebbens et~al.,
\newblock EPL {\bf 106}, 58003 (2014).

\bibitem{simmchen14a}
J.~Simmchen et~al.,
\newblock RSC Adv. {\bf 4}, 20334 (2014).

\bibitem{ebbens10a}
S.~J. Ebbens and J.~R. Howse,
\newblock Soft Matter {\bf 6}, 726 (2010).

\bibitem{hong10a}
Y.~Hong, D.~Velegol, N.~Chaturvedi, and A.~Sen,
\newblock Phys. Chem. Chem. Phys. {\bf 12}, 1423 (2010).

\bibitem{sengupta12a}
S.~Sengupta, M.~E. Ibele, and A.~Sen,
\newblock Angew. Chem. Int. Ed. {\bf 51}, 8434 (2012).

\bibitem{wang13a}
W.~Wang, W.~Duan, S.~Ahmed, T.~E. Mallouk, and A.~Sen,
\newblock Nano Today {\bf 8}, 531 (2013).

\bibitem{sanchez15a}
S.~S{\'a}nchez, L.~Soler, and J.~Katuri,
\newblock Angew. Chem. Int. Ed. {\bf 54}, 1414 (2015).

\bibitem{cates12a}
M.~Cates,
\newblock Rep. Prog.Phys. {\bf 75}, 042601 (2012).

\bibitem{cates15a}
M.~Cates and J.~Tailleur,
\newblock Ann. Rev. Cond. Mat. Phys. {\bf 6}, 219 (2015).

\bibitem{Sokolov07}
A.~Sokolov, I.~Aranson, J.~Kessler, and R.~Goldstein,
\newblock Phys. Rev. Lett. {\bf 98}, 158102 (2007).

\bibitem{Schwarz-Linek12}
J.~Schwarz-Linek et~al.,
\newblock Proc. Nat. Acad. Sci. {\bf 109}, 4052 (2012).

\bibitem{Reufer14}
M.~Reufer et~al.,
\newblock Biophys. J. {\bf 106}, 37 (2014).

\bibitem{Polin09}
M.~Polin, I.~Tuval, K.~Drescher, J.~Gollub, and R.~Goldstein,
\newblock Science {\bf 325}, 487 (2009).

\bibitem{Geyer13}
V.~Geyer, F.~J\"{u}licher, J.~Howard, and B.~Friedrich,
\newblock Proc. Nat. Acad. Sci. {\bf 110}, 18058 (2013).

\bibitem{Woolley03}
D.~Woolley,
\newblock Reproduction {\bf 126}, 259 (2003).

\bibitem{Riedel05}
I.~Riedel, K.~Kruse, and J.~Howard,
\newblock Science {\bf 309}, 300 (2005).

\bibitem{Ma14}
R.~Ma, G.~Klindt, I.~Riedel-Kruse, F.~J\"{u}licher, and B.~Friedrich,
\newblock Phys. Rev. Lett. {\bf 113}, 048101 (2014).

\bibitem{brown15a-pre}
A.~T. Brown, W.~C.~K. Poon, C.~Holm, and J.~de~Graaf,
\newblock arXiv {\bf 1512.01778}, 1 (2015).

\bibitem{wang14d}
S.~Wang and N.~Wu,
\newblock Langmuir {\bf 30}, 3477 (2014).

\bibitem{Solovev09}
A.~A. Solovev, Y.~Mei, E.~Berm{\'u}dez~Ure{\~n}a, G.~Huang, and O.~G. Schmidt,
\newblock Small {\bf 5}, 1688 (2009).

\bibitem{Mei11}
Y.~Mei, A.~A. Solovev, S.~Sanchez, and O.~G. Schmidt,
\newblock Chem. Soc. Rev. {\bf 40}, 2109 (2011).

\bibitem{kummel13}
F.~K{\"u}mmel et~al.,
\newblock Phys. Rev. Lett. {\bf 110}, 198302 (2013).

\bibitem{tenhagen14}
B.~Ten~Hagen et~al.,
\newblock Nat. Commun. {\bf 5}, 4829 (2014).

\bibitem{nijs2013}
B.~de~Nijs, A.~van Blaaderen, J.~Roeland, and J.~van Hest,
\newblock Nanoscale {\bf 5}, 1315 (2013).

\bibitem{vicsek95}
T.~Vicsek, A.~Czir{\'o}k, E.~Ben-Jacob, I.~Cohen, and O.~Shochet,
\newblock Phys. Rev. Lett. {\bf 75}, 1226 (1995).

\bibitem{zheng13}
X.~Zheng et~al.,
\newblock Phys. Rev. E {\bf 88}, 032304 (2013).

\bibitem{Stenhammar13}
J.~Stenhammar, A.~Tiribocchi, R.~Allen, D.~Marenduzzo, and M.~Cates,
\newblock Phys. Rev. Lett. {\bf 111}, 145702 (2013).

\bibitem{Redner13}
G.~Redner, A.~Baskaran, and M.~Hagan,
\newblock Phys. Rev. E {\bf 88}, 012305 (2013).

\bibitem{Thakur12}
S.~Thakur and R.~Kapral,
\newblock Phys. Rev. E {\bf 85}, 026121 (2012).

\bibitem{gonnella14}
G.~Gonnella, A.~Lamura, and A.~Suma,
\newblock Int. J. Mod. Phys. C {\bf 25}, 1441004 (2014).

\bibitem{Cugliandolo15}
L.~Cugliandolo, G.~Gonnella, and A.~Suma,
\newblock Phys. Rev. E {\bf 91}, 062124 (2015).

\bibitem{tung16}
C.~Tung, J.~Harder, C.~Valeriani, and A.~Cacciuto,
\newblock Soft Matter {\bf 12}, 555 (2016).

\bibitem{Peruani06}
F.~Peruani, A.~Deutsch, and M.~B{\"a}r,
\newblock Phys. Rev. E {\bf 74}, 030904 (2006).

\bibitem{wensink08}
H.~H. Wensink and H.~L{\"o}wen,
\newblock Phys. Rev. E {\bf 78}, 031409 (2008).

\bibitem{Ginelli10}
F.~Ginelli, F.~Peruani, M.~B{\"a}r, and H.~Chat{\'e},
\newblock Phys. Rev. Lett. {\bf 104}, 184502 (2010).

\bibitem{Yang10}
Y.~Yang, V.~Marceau, and G.~Gompper,
\newblock Phys. Rev. E {\bf 82}, 031904 (2010).

\bibitem{Abkenar13}
M.~Abkenar, K.~Marx, T.~Auth, and G.~Gompper,
\newblock Phys. Rev. E {\bf 88}, 062314 (2013).

\bibitem{isele15}
R.~Isele-Holder, J.~Elgeti, and G.~Gompper,
\newblock Soft Matter {\bf 11}, 7181 (2015).

\bibitem{hansen90}
J.-P. Hansen and I.~McDonald,
\newblock {\em Theory of simple liquids},
\newblock Elsevier, 1990.

\bibitem{frenkel88}
D.~Frenkel, H.~Lekkerkerker, and A.~Stroobants,
\newblock Nature {\bf 332}, 822 (1988).

\bibitem{bolhuis97}
P.~Bolhuis and D.~Frenkel,
\newblock J. Chem. Phys. {\bf 106}, 666 (1997).

\bibitem{wensink12}
H.~Wensink and H.~L{\"o}wen,
\newblock J. Phys.: Cond. Mat. {\bf 24}, 464130 (2012).

\bibitem{wensink14}
H.~Wensink, V.~Kantsler, R.~Goldstein, and J.~Dunkel,
\newblock Phys. Rev. E {\bf 89}, 010302 (2014).

\bibitem{mallory16}
S.~Mallory and A.~Cacciuto,
\newblock arxiv {\bf 1606.05362}, 1 (2016).

\bibitem{prymidis16}
V.~Prymidis, S.~Samin, and L.~Filion,
\newblock Soft Matter  (2016).

\bibitem{alizadehrad15}
D.~Alizadehrad, T.~Kr{\"u}ger, M.~Engstler, and H.~Stark,
\newblock PLoS Comput. Biol. {\bf 11}, e1003967 (2015).

\bibitem{Elgeti10}
J.~Elgeti, U.~Kaupp, and G.~Gompper,
\newblock Biophys. J. {\bf 99}, 1018 (2010).

\bibitem{Hu15}
J.~Hu, M.~Yang, G.~Gompper, and R.~Winkler,
\newblock Soft Matter {\bf 11}, 7867 (2015).

\bibitem{kaiser13}
A.~Kaiser, K.~Popowa, H.~H. Wensink, and H.~L{\"o}wen,
\newblock Phys. Rev. E {\bf 88}, 022311 (2013).

\bibitem{Note1}
Note that our equation of motion is 3D and we therefore must employ quaternions
  to specify 3D rotation.

\bibitem{martys99}
N.~Martys and R.~Mountain,
\newblock Phys. Rev. E {\bf 59}, 3733 (1999).

\bibitem{ahlrichs99}
P.~Ahlrichs and B.~D{\"u}nweg,
\newblock J. Chem. Phys. {\bf 111}, 8225 (1999).

\bibitem{lobaskin04}
V.~Lobaskin and B.~D\"unweg,
\newblock New J. Phys. {\bf 6}, 54 (2004).

\bibitem{roehm14}
D.~Roehm, S.~Kesselheim, and A.~Arnold,
\newblock Soft Matter {\bf 10}, 5503 (2014).

\bibitem{fischer15}
L.~Fischer, T.~Peter, C.~Holm, and J.~de~Graaf,
\newblock J. Chem. Phys. {\bf 143}, 084107 (2015).

\bibitem{degraaf15b}
J.~de~Graaf, T.~Peter, L.~Fischer, and C.~Holm,
\newblock J. Chem. Phys. {\bf 143}, 084108 (2015).

\bibitem{degraaf16a}
J.~de~Graaf et~al.,
\newblock J. Chem. Phys. {\bf 144}, 134106 (2016).

\bibitem{degraaf16b}
J.~de~Graaf et~al.,
\newblock Soft Matter {\bf accepted},  (2016).

\bibitem{limbach06a}
H.~J. Limbach, A.~Arnold, B.~A. Mann, and C.~Holm,
\newblock Comp. Phys. Comm. {\bf 174}, 704 (2006).

\bibitem{arnold13a}
A.~Arnold et~al.,
\newblock {ESPResSo 3.1 --- Molecular Dynamics Software for Coarse-Grained
  Models},
\newblock in {\em Meshfree Methods for Partial Differential Equations {VI}},
  edited by M.~Griebel and M.~A. Schweitzer, volume~89 of {\em Lecture Notes in
  Computational Science and Engineering}, page~1, Springer, 2013.

\bibitem{ortega03}
A.~Ortega and J.~de~la Torre,
\newblock J. Chem. Phys. {\bf 119}, 9914 (2003).

\bibitem{mermin66}
N.~Mermin and H.~Wagner,
\newblock Phys. Rev. Lett. {\bf 17}, 1133 (1966).

\bibitem{kraft10}
D.~Kraft et~al.,
\newblock J. Phys. Chem. B {\bf 115}, 7175 (2010).

\bibitem{kraft13}
D.~Kraft et~al.,
\newblock Phys. Rev. E {\bf 88}, 050301 (2013).

\bibitem{lin15}
N.~Lin et~al.,
\newblock Phys. Rev. Lett. {\bf 115}, 228304 (2015).

\bibitem{deseigne10}
J.~Deseigne, O.~Dauchot, and H.~Chat{\'e},
\newblock Phys. Rev. Lett. {\bf 105}, 098001 (2010).

\bibitem{kumar14}
N.~Kumar, H.~Soni, S.~Ramaswamy, and A.~Sood,
\newblock Nat. Commun. {\bf 5}, 4688 (2014).

\end{thebibliography}
\end{document}